\documentclass[[12pt]{article}
\usepackage{verbatim}
\usepackage[english]{babel}
\usepackage{pifont}
\usepackage{amsmath}
\usepackage{graphicx}
\usepackage{natbib}
 \usepackage{mathptmx}      
\usepackage{latexsym}

\begin{document}

\title{Thermodynamics of long-run economic innovation
and growth}

\author
{Timothy J. Garrett }



\date{}

\maketitle
\begin{abstract}
This article derives prognostic expressions for the evolution of globally
aggregated economic wealth, productivity, inflation, technological
change, innovation and growth. The approach is to treat civilization
as an open, non-equilibrium thermodynamic system that dissipates energy
and diffuses matter in order to sustain existing circulations and
to further its material growth. Appealing to a prior result that established
a fixed relationship between a very general representation of global
economic wealth and rates of global primary energy consumption, physically
derived expressions for economic quantities follow. The analysis suggests
that wealth can be expressed in terms of the length density of civilization's
networks and the availability of energy resources. Rates of return
on wealth are accelerated by energy reserve discovery, improvements
to human and infrastructure longevity, and a more common culture,
or a lowering of the amount of energy required to diffuse raw materials
into civilization's bulk. According to a logistic equation, rates
of return are slowed by past growth, and if rates of return approach
zero, such ``slowing down'' makes civilization fragile with respect
to externally imposed network decay. If past technological change
has been especially rapid, then civilization is particularly vulnerable
to newly unfavorable conditions that might force a switch into a mode
of accelerating collapse. 
\end{abstract}

\section{Introduction}

Like other natural systems, civilization is composed of matter, and
its internal circulations are maintained through a dissipation of
potential energy. Oil, coal, and other fuels ``heat'' civilization
to raise the potential of its internal components. Frictional, resistive,
radiative, and viscous forces return the potential of civilization
to its initial state, ready for the next cycle of energy consumption.
Burning coal at a power station raises an electrical potential or
voltage which then allows for a down-voltage electrical flow; the
potential energy is dissipated at some point between the power station
and the appliance; because what the appliance does is useful, a human
demand is sustained for more coal to burn. Similarly, energy is dissipated
as cars burn gasoline to propel vehicles to and from desirable destinations.
Or, people consume food to maintain the circulations of their internal
cardiovascular, respiratory, and nervous systems while dissipating
heat and renewing their hunger. 

Such cycles are fairly fast; at least the longest might be the annual
periodicities that are tied to agriculture. This paper provides a
framework for the slower evolution of civilization over timescales
where such rapid cyclical behavior tends to average out. Instead,
the perspective is that material growth and decay of civilization
networks is driven by a long-run imbalance between energy consumption
and dissipation. 

The approach that is followed here builds upon a more general treatment
for the evolution of natural systems that has been outlined previously
in \citet{Garrettmodes2012}, which starts from first thermodynamic
principles in order to develop a fairly general expression for the
spontaneous emergence of natural systems. From this point, analytical
expressions are provided for economic growth that can be expressed
in units of currency. These are then presented in a form that can
be evaluated against economic statistics for past behavior and be
used to provide physically constrained scenarios for the future.

\section{\label{sec:Energetic-and-material-flows}Energetic and material flows
to systems}

\begin{figure}[h]
\includegraphics[width=4.5in]{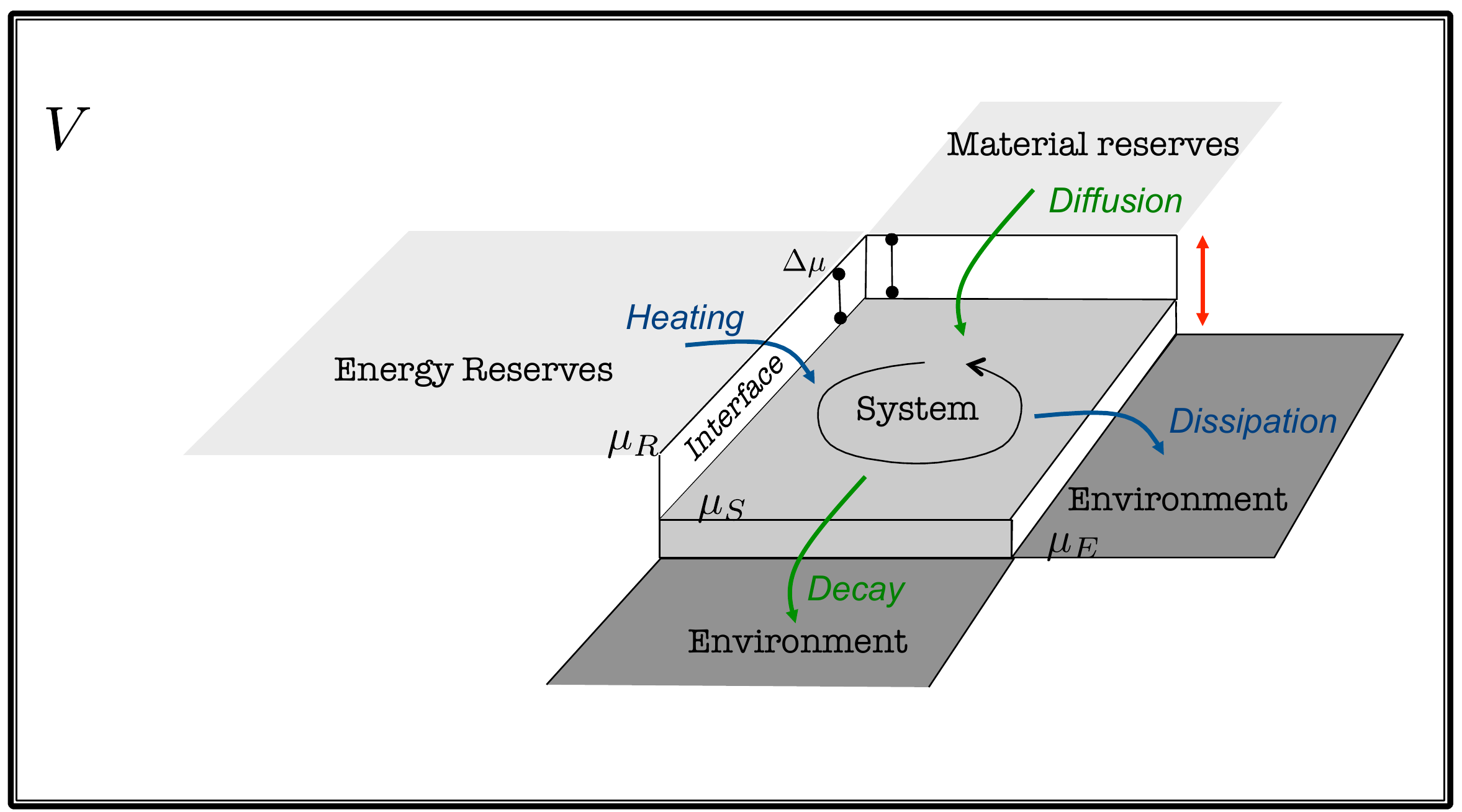}

\caption{\label{fig:equilibrium}\emph{Schematic for the thermodynamics of
an open system within a fixed volume $V$. Energy reserves, the system,
and the environment lie along distinct constant potential surfaces
$\mu_{R}$, $\mu_{S}$, and $\mu_{E}$. Internal material circulations
within the system are sustained by heating and dissipation of energy
that is coupled to a material flow of diffusion and decay. The level
$\mu_{S}$ is a time-averaged potential. Over shorter time-scales,
the legs of a heat-engine cycle would show the system rising up and
down between $\mu_{E}$ and $\mu_{R}$ in response to heating and
dissipation, as shown by the red arrow, allowing for material diffusion
to the system and decay from the system. If flows are in balance then
the system is at equilibrium and it does not grow. }}
\end{figure}
The universe is a continuum of matter and potential energy in space.
Local gradients drive thermodynamic flows that redistribute matter
and energy over time. In the sciences, we invoke the existence of
some ``system'' or ``particle'' from within this continuum, requiring
as a first step that we define some discrete contrast between the
system and its surroundings as shown in Fig. \ref{fig:equilibrium}.
This discrete contrast can be approximated as an interfacial jump
in potential energy $\Delta\mu$ between the system potential $\mu_{S}$
and some higher level $\mu_{R}$; or, $\Delta\mu=\mu_{S}-\mu_{E}$
with respect to a lower level $\mu_{E}$. Matter that lies along the
higher potential $\mu_{R}$ has a higher temperature and/or pressure,
so it can be viewed as a ``reserve'' for downhill flows that ``pour''
into the system. Flows also ``drain'' to the lower potential environment
lying along the potential surface $\mu_{E}$. 

Viewed from a strictly thermodynamic perspective, any system that
is defined by a constant potential implicitly lies along a smooth
surface within which there is no resolved internal contrast, i.e.,
one where there is a fixed potential energy per unit matter $\mu_{S}$
and no internal gradients. This specific potential represents the
time-integrated quantity of work that has been required to displace
each unit of matter within the surface through an arbitrary set of
force-fields that point in the opposite direction of the potential
vector $\mu$: for example, the gravitational potential per block
in a pyramid is determined by the product of the downward gravitational
force on each block and its height. 

Although internal gradients and circulations are not resolved within
a constant potential surface, the presence of the continuum requires
that they exist nonetheless. For example, when a bathtub is filled,
internal gradients force the water to slosh from side to side. While,
the short timescale of these small waves might be of interest to a
child, a typical adult cares only about the time-averaged water level
of the bathtub as a whole, and that it gradually rises as the water
pours in. The definition of what counts as a ``system'' is only
a matter of perspective. It depends on what timescale is of most interest
to the observer looking at the system's variability. As a general
rule, however, coarse spatial resolution corresponds with coarse time
resolution \citep[e.g.,][]{Blois2013}.

The total energy of a system, or its enthalpy $H_{S}$, can be expressed
as a product of the amount of matter in the system $N_{S}$ and the
specific enthalpy given by 
\begin{equation}
e_{S}^{tot}=\left(\frac{\partial H_{S}}{\partial N_{S}}\right)_{\mu_{S}}\label{eq:e_s_tot}
\end{equation}
The specific enthalpy can be decomposed into the product of the total
number of independent degrees of freedom $\nu$ in the system and
the oscillatory energy per independent degree of freedom $e_{S}$%
\footnote{For example, nitrogen gas at atmospheric temperatures and pressures
has a specific enthalpy that is the product of the specific heat at
constant pressure $c_{p}$ and the system temperature $T_{S}$, or
$e_{S}^{tot}=c_{p}T_{S}$. The specific enthalpy can be decomposed
into $\nu=7$ degrees of freedom. The internal energy has three translational
degrees and two rotational degrees. Plus there are an additional two
effective degrees that are associated with the pressure energy within
a volume. Each degree of freedom has a time-averaged kinetic energy
equal to $kT_{S}/2$ where $k$ is the Boltzmann constant. %
} 
\begin{equation}
e_{S}^{tot}=\nu e_{S}\label{eq:e_s_tot_nu}
\end{equation}
The quantity $e_{S}$ represents the circulatory energy per degree
of freedom per unit matter. Thus, 
\begin{equation}
H_{S}\left(\mu_{S}\right)=N_{S}e_{S}^{tot}=\nu N_{S}e_{S}\label{eq:H_S}
\end{equation}

Conservation of energy considerations dictate that enthalpy is the
energetic quantity that rises when there is net heating of the system
at a constant pressure \citep{Zemanksy1997}, i.e. 
\begin{equation}
\left(\frac{\partial H_{S}}{\partial t}\right)_{p}=\left(\frac{\partial Q^{net}}{\partial t}\right)_{p}\label{eq:dH_S_dt_p}
\end{equation}
and that net heating of the system is a balance between a supply of
energy to the system at rate $a$ and a dissipation at rate $d$ 
\begin{equation}
\left(\frac{\partial Q^{net}}{\partial t}\right)_{p}=a-d\label{eq:dQdt}
\end{equation}
The Second Law requires that dissipation is to some lower potential
and that heating drains some higher potential reserve of enthalpy.
Not all enthalpy in the reserve $H_{R}$ is necessarily \emph{available}
to the system. For example, unless the temperature of the system is
raised to extremely high levels, the nuclear enthalpy of a reserve
$H_{R}=mc^{2}$ might normally be inaccessible. Thus, available enthalpy
is distinguished here by the symbol $\Delta H_{R}$.

Heating is coupled to material flows in what can be idealized as a
four step cycle termed a ``heat engine'', whose circulation is shown
by the red arrow in Fig. \ref{fig:equilibrium}. A system that is
initially in equilibrium with the environment at level $\mu_{E}$
is heated, which raises the potential level of the system $\mu_{S}$
an amount $2\Delta\mu$ to level $\mu_{R}$ with a timescale of $\tau_{heat}\sim2\Delta\mu/a$.
It is at this point that the surface $\mu_{S}$ comes into diffusive
equilibrium with respect to external sources of raw materials, allowing
for a material flow to the system \citep{kittel1980thermal}%
\footnote{A well-known expression of this physics is the Gibbs-Duhem equation
\citep{Zemanksy1997}.%
}. There is then cooling through dissipation of heat to the environment
with timescale $\tau_{diss}\sim2\Delta\mu/d$, which brings the system
back into diffusive equilibrium with surface $\mu_{E}$, allowing
for material decay. 

How the thermodynamics is treated depends on whether the timescale
of interest is short or long compared to $\tau_{heat}$.

\subsection{Systems in material equilibrium}

Over time scales much shorter than $\tau_{heat}$, the legs of the
heat engine are resolved, so that the amount of matter in a system
$N_{S}$ would appear to change sufficiently slowly that it could
be considered to be fixed. In this case, the response to net heating
would be that the specific enthalpy per unit matter rises at rate
\begin{equation}
\left(\frac{\partial e_{S}^{tot}}{\partial t}\right)_{p,N_{S}}=\frac{1}{N_{S}}\left(\frac{\partial Q^{net}}{\partial t}\right)_{p,N_{S}}\label{eq:des_totdt}
\end{equation}
For the example that heating is a response to radiative flux convergence,
then it may be that the temperature rises according to:
\begin{equation}
c_{p}\left(\frac{\partial T}{\partial t}\right)_{p,N_{S}}=\frac{1}{N_{S}}\left(\frac{\partial Q^{net}}{\partial t}\right)_{p,N_{S}}\label{eq:dTdt}
\end{equation}
where $c_{p}$ is the specific heat of the substance at constant pressure
and $\partial Q^{net}/\partial t$ is the radiative heating. In a
materially closed system, the response to net heating is for the temperature
to rise.

In the atmospheric sciences, Eq. \ref{eq:dTdt} expresses how radiative
heating is the driving force behind weather \citep{Liou-book}. At
timescales longer than $\tau_{heat}$, however, the establishment
of a temperature gradient ultimately leads to a material flow that
we call the wind.

\subsection{\label{sub:Systems-in-material-disequilibrium}Systems in material
disequilibrium}

Over timescales much longer than $\tau_{heat}$, the legs of the
heat engine are not resolved. Instead, because the heat engine cycles
are much faster than the timescales of interest, one only views some
average level of $\mu_{S}$ that lies in between the points of maximum
and minimum potential energy, $\mu_{R}$ and $\mu_{E}$ (Fig. \ref{fig:equilibrium}). 

\begin{figure}[h]
\noindent \begin{centering}
\includegraphics[width=4.5in]{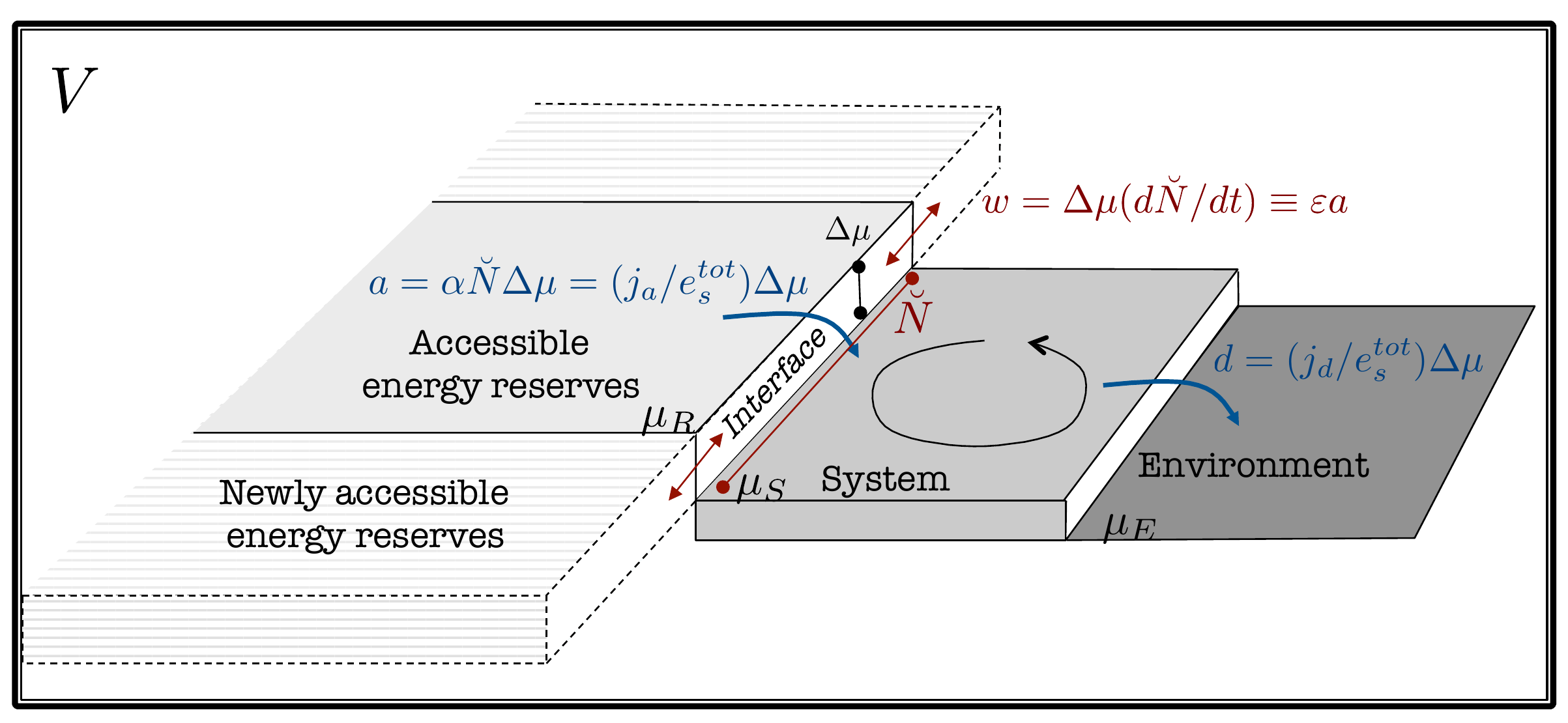}
\par\end{centering}

\caption{\label{fig:diffusion}\emph{Schematic for the thermodynamic evolution
of a system within a constant volume $V$. Energy reserves, the system,
and the environment lie along distinct constant potential surfaces
$\mu_{R}$, $\mu_{S}$, and $\mu_{E}$. The size of an interface $\breve{N}\Delta\mu$
between surfaces determines the rate of heating $a$ and the speed
of downhill material flow $j_{a}$. The system grows or shrinks according
to a net material flux convergence $j_{a}-j_{d}$ along $\mu_{S}$.
System growth is related to expansion work $w$ that is done to grow
the interface, extending the system's access to previously inaccessible
energy reserves. The efficiency of work is determined by $\varepsilon=w/a$.}}
\end{figure}
In this case, energetic and material flows appear to be coupled. An
illustration of this coupling is shown in Fig. \ref{fig:diffusion},
which recasts Fig. \ref{fig:equilibrium} in terms of a single co-ordinate.
Where there is a disequilibrium, material convergence along a surface
of constant potential $\mu_{S}$ corresponds with growth of the system
enthalpy at rate
\begin{equation}
\left(\frac{\partial H_{S}}{\partial t}\right)_{\mu_{S}}=\left(\frac{\partial Q^{net}}{\partial t}\right)_{\mu_{S}}=e_{S}^{tot}\left(\frac{\partial N_{S}}{\partial t}\right)_{\mu_{S}}\label{eq:dH_Sdt_mu}
\end{equation}
so that from Eq. \ref{eq:dQdt}, the bulk grows at rate 
\begin{eqnarray}
\left(\frac{\partial N_{S}}{\partial t}\right)_{\mu_{S}} & = & \frac{\left(\partial Q^{net}/\partial t\right)_{\mu_{S}}}{e_{S}^{tot}}\label{eq:dNsdt_mu}\\
 & = & \frac{a-d}{e_{S}^{tot}}\nonumber 
\end{eqnarray}
If there is zero time-averaged net heating, then $\left\langle \left(\partial Q^{net}/\partial t\right)_{\mu_{S}}\right\rangle =0$
because $\left\langle a\right\rangle =\left\langle d\right\rangle $,
in which case the size of the system $N_{S}$ does not change. Like
water pouring into and draining from a bath tub at equal rates, circulations
within the system maintain a steady-state%
\footnote{For the case of zero net heating, there is nonetheless an increase
in global entropy even though local entropy production $\left(\partial Q^{net}/\partial t\right)_{\mu}/\mu=0$
. A continuous flow from high to low potential requires increasing
global entropy $\sum_{\mu}\left(\partial Q^{net}/\partial t\right)_{\mu}/\mu$
because there is global redistribution of matter to low values of
$\mu$. %
}

Material growth occurs for the non-equilibrium condition that energy
consumption exceeds dissipation, in which case $\left\langle \left(\partial Q^{net}/\partial t\right)_{\mu_{S}}\right\rangle >0$.
In this case, there is a net convergence of matter along the potential
surface $\mu_{S}$ at rate $j^{net}$. Material flows into civilization
at rate $j_{a}$ and out of civilization at the decay rate $j_{d}$
form a balance defined by 
\begin{eqnarray}
j^{net}=\left(\frac{\partial N_{S}}{\partial t}\right)_{\mu_{S}} & = & j_{a}-j_{d}\label{eq:jnet}
\end{eqnarray}
so that the timescale for growth of the system is $\tau_{growth}\sim N_{S}/j^{net}$.
Combined with Eq. \ref{eq:dNsdt_mu}, this implies that 
\begin{eqnarray}
j_{a} & = & a/e_{S}^{tot}\label{eq:ja}\\
j_{d} & = & d/e_{S}^{tot}\label{eq:jd}\\
j^{net} & = & \frac{a-d}{e_{S}^{tot}}\label{eq:jnet-1}
\end{eqnarray}
A straightforward and familiar example of this physics is what happens
when we boil a pot of water. Once the water reaches the boiling point,
the temperature of the water is maintained at a constant 100$^{\circ}$
C, and the energy input from the stove goes into turning liquid water
into bubbles. Setting aside the energetics of forming the bubble surface,
and assuming the pot is well insulated, the energy input that is required
to vaporize a single liquid water molecule is $e_{S}^{tot}=l_{v}$
where $l_{v}$ is the latent heat of evaporation at boiling. Thus,
vapor molecules contained in the bubbles are created at a rate that
is proportional to the rate of energetic input: $j_{a}=a/e_{S}^{tot}=a/l_{v}$. 

Heating creates an internal circulation of bubbles that we call a
boil. When bubbles rise to the surface, molecules escape the fluid
at rate $j_{d}$, and there is an associated evaporative cooling of
the water at rate $d=j_{d}e_{S}^{tot}=j_{d}l_{v}$. With a steady
simmer, a constant vapor concentration $N_{S}$ is maintained within
the pot because heating equals cooling. In this case, from Eq. \ref{eq:jnet-1},
$j_{a}\simeq j_{d}$ and $j^{net}=0$. 

If the output from the heating element is suddenly raised to high,
then there is a non-equilibrium adjustment period of $\tau_{growth}\sim N_{S}/(j_{a}-j_{d})$
during which heating temporarily exceeds dissipation and bubble production
at the bottom of the pot $j_{a}$ exceeds bubble popping at its top
$j_{d}$. The size and number of vapor bubbles in the water increases,
and a new stasis is attained only when evaporative cooling $d$ rises
to come into equilibrium with the element heating $a$. At this point,
the pot has gone from a simmer to a rolling boil.

\subsection{Gradients and flows }

As shown in Fig. \ref{fig:diffusion}, a material flow at rate $j$
can seen as a diffusion of matter downhill as it flows across a material
interface. The interface between the system and its higher potential
reservoirs can be defined by a potential step with a rise $\Delta\mu=\mu_{R}-\mu_{S}$
and an orthogonal quantity of material that lies along the interface
$\breve{N}$. The total energy required to grow the interface is the
product of these two quantities: i.e., $\Delta G=\breve{N}\Delta\mu$.
Because the gradient enables flows, there is a proportional consumption
of available potential energy $\Delta H_{R}$ at rate 
\begin{eqnarray}
a & = & \alpha\Delta G=\alpha\breve{N}\Delta\mu\label{eq:a_alphanhat}
\end{eqnarray}
where $\alpha$ is a rate coefficient with units of inverse time.
The quantity $\Delta G=\breve{N}\Delta\mu$ in Eq. \ref{eq:a_alphanhat}
differs from the available enthalpy $\Delta H_{R}=N_{R}\Delta\mu$.
The available enthalpy is a reserve of energy, but it is $\Delta G$
that is associated with the gradient that drives flows across an interface. 

From Eqs. \ref{eq:jnet} and \ref{eq:ja}, energy consumption is coupled
to a material flux $j_{a}=\left(\partial N_{S}/\partial t\right)_{\mu_{S}}$.
Thus, from Eq. \ref{eq:a_alphanhat}: 
\begin{eqnarray}
j_{a} & = & \alpha\breve{N}\Delta\mu/e_{S}^{tot}\label{eq:jint_sigma_a}
\end{eqnarray}
The magnitude of the interface $\breve{N}$ reflects the respective
sizes of the two components it separates. In general, when there is
a diffusive flow to a system, $\breve{N}$ is proportional to a product
of the available enthalpy within a high potential energy ``reservoir''
$\Delta H_{R}=N_{R}\Delta\mu$ and the size of the system $N_{S}$
taken to a one third power \citep{Garrettmodes2012}, or that 
\begin{equation}
\breve{N}=kN_{S}^{1/3}N_{R}\label{eq:Nbreve}
\end{equation}
where a dimensionless coefficient $k$ is related to the object shape
\footnote{For a system that is spherical with respect to its reserves then $k=\left(48\pi^{2}\right)^{1/3}$
\citep{Garrettmodes2012}%
}. 

At first glance, one might guess that the system interface should
be proportional $N_{S}N_{R}$ instead, since both the size of the
system and the size of the reserve are what drive flows between the
two. A system's size is proportional to its volume $V_{S}=N_{S}/n_{S}$,
where $N_{S}$ is the number of elements in the system and $n_{S}$
is the internal density; $V_{S}$ and $N_{S}$ are proportional to
a dimension of length cubed, or volume. However, flows to a system
are not determined by a volume. Rather, flows are down a linear gradient
that lies normal to a surface. The surface area has dimensions of
length squared or $N_{S}^{2/3}$, and the linear gradient has dimensions
of inverse length or $N_{S}^{-1/3}$. Both factors control the flow
rate, and their product yields a one third power or a length dimension:
$N_{S}^{2/3}\times N_{S}^{-1/3}=N_{S}^{1/3}$. 

In any case, if it were assumed that $\breve{N}$ is proportional
to the product $N_{S}N_{R}$, then the implication would be that wholes
are interacting with wholes. A perfect mixture of the system and its
reserve, even if possible (which it is not), would make it impossible
to resolve flows between $N_{S}$ and $N_{R}$: the two components
would be indistinguishable. Finally, assuming a unity exponent for
$N_{S}$ removes any element of persistence or memory from rates of
system growth, as will be shown below. Unphysically, it would divorce
what happens in the present from what has happened in the past. 

Since $\Delta H_{R}=N_{R}\Delta\mu$ , Eqs. \ref{eq:a_alphanhat}
and \ref{eq:jint_sigma_a} for energy dissipation and material flows
can now be expressed as 
\begin{eqnarray}
j & = & \alpha kN_{S}^{1/3}\Delta H_{R}/e_{S}^{tot}\label{eq:flowsNbreve_general}\\
a & = & \alpha kN_{S}^{1/3}\Delta H_{R}
\end{eqnarray}
In \citet{Garrettmodes2012} it was shown that the quantity $\alpha kN_{S}^{1/3}$
can be expressed in an equivalent fashion in terms of a length density
times a diffusivity $\Lambda\mathcal{D}$, where the length density
is analogous to the electrostatic capacitance within a volume and
the diffusivity has dimensions of area per time%
\footnote{A very simple example of this physics is the diffusional growth of
a spherical cloud droplet of radius $r$ through the condensation
of water vapor, where $j=4\pi r\mathcal{D}N_{R}/V$ and $N_{R}/V$
is equivalent to the excess vapor density relative to saturation.
In this case $\alpha kN_{S}^{1/3}\mathcal{D}=\Lambda\mathcal{D}=4\pi r\mathcal{D}/V$.
Note that a length dimension is what determines flows, insofar as
it is coupled to available reserves of potential energy. For more
dendritic structures like snowflakes, there is no clearly definable
``radius'', yet it is still a length dimension $\Lambda$ or ``capacitance''
that drives diffusive growth \citep{PruppacherKlett1997}.%
}. Thus, the flow and dissipation equations can be alternatively expressed
as 
\begin{eqnarray}
j & = & \mathcal{D}\Lambda\Delta H_{R}/e_{S}^{tot}\label{eq:flowsNbreve}\\
a & = & \mathcal{D}\Lambda\Delta H_{R}\label{eq:aflowslambda}
\end{eqnarray}
The rate of material flows is proportional to a rate of energy dissipation
$a$, which in turn is proportional to some measure of the length
density within the system $\Lambda$ or its accumulated size $N_{S}$
to a one third power, and the number of potential energy units in
the reserve $N_{R}=\Delta H_{R}/\Delta\mu$. The final component is
$e_{S}^{tot}$, which expresses the amount of energy that must be
dissipated to enable each unit of material flow towards the system.

\subsection{Efficiency and growth}

As described above, a system grows if there is net heating that drives
an imbalance between diffusive material flows (Eqs. \ref{eq:jnet}
and \ref{eq:jnet-1}), so that the size of the system $N_{S}$ and
interface with energy reserves $\Delta G=\breve{N}\Delta\mu$ evolve
over time. 

Taking the approach that the resolved rise of the interface $\Delta\mu$
is fixed, then flows evolve as the magnitude of the ``step'' $\breve{N}\Delta\mu$
grows laterally (Fig. \ref{fig:diffusion}). Here, this material expansion
or ``stretching'' of the interface $\breve{N}$ and the potential
difference $\Delta G$ is termed ``work'' $w$, where:
\begin{equation}
w=\left(\frac{\partial\Delta G}{\partial t}\right)_{\mu_{R},\mu_{S}}=\left(\frac{\partial\breve{N}}{\partial t}\right)_{\mu_{R},\mu_{S}}\Delta\mu\label{eq:work}
\end{equation}
The efficiency of converting heating to a rate of doing work is normally
defined by the ratio 
\begin{equation}
\varepsilon=\frac{w}{a}\label{eq:efficiency}
\end{equation}
Here, efficiency can be either positive or negative depending on whether
the interface is shrinking or growing in response to heating, and
therefore on the sign of $w$ (Eq. \ref{eq:work}). 

From Eq. \ref{eq:work}, the relative growth rate of the interface
can be defined by 
\begin{equation}
\eta=\frac{w}{\Delta G}=\frac{d\ln\Delta G}{dt}=\frac{d\ln\breve{N}}{dt}\label{eq:eta_logarithm-1}
\end{equation}
where $\eta$ has units of inverse time. In other words, $1/\eta$
is the characteristic time for exponential growth of $\Delta G$ and
$\breve{N}$. 

Since, from Eqs. \ref{eq:work} and \ref{eq:efficiency}, $w=d\Delta G/dt=\varepsilon a$
and from Eq. \ref{eq:a_alphanhat}, $a=\alpha\Delta G$, it follows
that the relationship between the growth rate $\eta$ and efficiency
$\varepsilon$ and heating $a$ is given by
\begin{eqnarray}
\eta & = & \alpha\varepsilon\label{eq:eta_logarithm}\\
 & = & \frac{d\ln a}{dt}
\end{eqnarray}
which has the advantage of expressing $\eta$ in terms of a measurable
flux $a$. So, efficient systems grow faster to consume more. For
the special case of pure exponential growth where $\eta$ is a constant,
then $a=a_{0}\exp\left(\eta t\right)$, but, more generally, nothing
is ever fixed in time: $\eta$ constantly changes as the interface
evolves, and it can even change sign if it shrinks. The growth rate
$\eta$ is positive if the efficiency $\varepsilon$ is greater than
zero meaning that the system is able to do net work on its surroundings
in response to heating (i.e. $d\ln\breve{N}/dt>0$). Otherwise, the
growth rate is negative and the system collapses (i.e. $\varepsilon<0$
and $d\ln\breve{N}/dt<0)$.

\subsection{Emergence, diminishing returns, and decay }

A pot of boiling pot of water has an external agency with its hand
on the energetic flow. ``Emergent systems'' might be characterized
by a spontaneous development of a structure. A way to view emergence
is through Fig. \ref{fig:diffusion}, where heating and dissipation
sustain internal circulations. If heating exceeds dissipation then
a net incorporation of matter into the system allows it to expand
into newly accessible energy reserves. The thermodynamic recipe for
emergence is that sufficient energy reserves exist to be ``discovered''
that the disequilibrium that drives growth can be sustained.

While emergent phenomena are ubiquitous in nature, they might be most
evident in living organisms who survive by eating, drinking and inhaling
a matrix of matter and potential energy, which is then diffused through
a linear of network of vascular structures. Consumption of the potential
energy in carbohydrates, proteins and fats sustains the organism and
facilitates an incorporation of water, chemicals, vitamins, minerals.
Meanwhile, heat is dissipated, and matter is lost, through radiation,
perspiration, exhalation, and excretion. The flow of raw materials
and the dissipation of potential energy are coupled within cardiovascular,
respiratory, gastro-intestinal and nervous networks. Over short timescales,
dissipation simply allows for further consumption. In the long-run
though, where consumption is in excess of dissipation, flows are out
of equilibrium, and the organism networks grow. The demand for energy
by the organism increases out of a requirement to sustain its growing
network length and the associated internal circulations. 

For a given availability of available energy supplies $\Delta H=N_{R}\Delta\mu$,
then from Eqs. \ref{eq:Nbreve} and \ref{eq:eta_logarithm-1} the
instantaneous growth rate is related to the system size $N_{S}$ or
its network length density $\Lambda$ through 
\begin{eqnarray}
\eta & = & \left(\frac{\partial\ln\breve{N}}{\partial t}\right)_{N_{R}}\label{eq:dnbrevedt}\\
 & = & \left(\frac{\partial\ln N_{S}^{1/3}}{\partial t}\right)_{N_{R}}\label{eq:dNsdt}\\
 & = & \left(\frac{\partial\ln\Lambda}{\partial t}\right)_{N_{R}}\label{eq:dlnLambda/dt}
\end{eqnarray}
If the rate of emergent growth $\eta$ is positive then a positive
feedback loop dominates and this length dimension grows exponentially
(i.e. $\Lambda=\Lambda_{0}\exp{{\left(\eta t\right)}}$).
Negative values of $\eta$ correspond with decay.

From Eq. \ref{eq:jnet} and \ref{eq:dNsdt}, the rate of emergent
growth can be related to rates of material consumption $j_{a}$ and
decay $j_{d}$ through: 
\begin{eqnarray}
\eta & = & \frac{1}{3N_{S}}\left(\frac{\partial N_{S}}{\partial t}\right)_{N_{R}}\label{eq:eta_approx}\\
 & = & \frac{1}{3}\frac{j_{a}-j_{d}}{\int_{0}^{t}\left(j_{a}-j_{d}\right)dt'}\\
 & = & \frac{1}{3}\frac{j^{net}}{\int_{0}^{t}j^{net}dt'}
\end{eqnarray}
Note that the timescale for growth of the system discussed earlier
$\tau_{growth}$ is related to the growth rate of flows through $\eta=3/\tau_{growth}$. 

A ``decay parameter'' $\delta$ can be defined as the rate of material
decay relative to the rate of material consumption: 
\begin{equation}
\delta=\frac{j_{d}}{j_{a}}\label{eq:delta}
\end{equation}
and, since the current system size is the time integral of past net
material flows, $N_{S}=\int_{0}^{t}j^{net}dt'$, it follows that the
rate of emergent growth is given by: 
\begin{eqnarray}
\eta & = & \frac{1}{3}\left(1-\delta\right)\frac{j_{a}}{N_{S}}\label{eq:eta_kappa_ja}\\
 & = & \frac{1}{3}\frac{\left(1-\delta\right)j_{a}}{\int_{0}^{t}\left(1-\delta\right)j_{a}dt'}
\end{eqnarray}

The final step is to account for the motive force for current flows
to the system, which is obtained by substituting Eq. \ref{eq:flowsNbreve_general}
into Eq. \ref{eq:eta_kappa_ja} to yield 
\begin{eqnarray}
\eta & = & \alpha k\left(1-\delta\right)\frac{N_{S}^{1/3}N_{R}\Delta\mu}{N_{S}e_{S}^{tot}}\label{eq:eta-past}\\
 & = & \alpha k\left(1-\delta\right)\frac{\Delta H_{R}}{N_{S}^{2/3}e_{S}^{tot}}
\end{eqnarray}
Eq. \ref{eq:eta-past} for emergent growth has seven parameters. Three
-- $\alpha$, $k$ and $\Delta\mu$ -- are considered as constants
in this treatment. So, current growth rates $\eta$ are determined
by the quantity of energy $\Delta H_{R}=N_{R}\Delta\mu$ that is available
to drive material flows to the system; the amount of energy $e_{S}^{tot}$
that must be dissipated to incorporate each unit of matter into the
system; the fraction $1-\delta$ of this new matter whose addition
is not offset by decay; and, crucially, past flows leading to the
current system size $N_{S}$: as a system gets bigger, there is a
natural propensity for its growth rate to slow with time. 

This last element leads to a ``law of diminishing returns'' and
introduces memory to emergent growth. Note that, had it been assumed
that flows were proportional to $N_{S}N_{R}\Delta\mu$ rather than
$N_{S}^{1/3}N_{R}\Delta\mu$ in Eq. \ref{eq:flowsNbreve_general},
then this dependence of current growth rates on past flows $\int_{0}^{t}j^{net}dt'$
would not be present -- the $N_{S}$ terms would have canceled in
Eq. \ref{eq:eta-past}. Clearly, this would be inconsistent with our
observations of emergent systems. Expressed logarithmically, large
objects tend to grow more slowly than small objects. And, the growth
of all emergent systems is somehow tied to their past. Systems are
built from matter that was accumulated during prior growth. ``Great
oaks from little acorns grow''.

\section{Thermodynamics of the growth of wealth }

Taken as a whole, civilization might be viewed as another example
of an emergent system that, like other living organisms, consumes
``food'' in the form of a matrix of matter and energy. The raw materials
include water, wood, cement, copper and steel. The potential energy
that is consumed is contained in fossil fuels, nuclear fuels, and
renewables. The linear networks within civilization are our roads,
shipping lanes, communication links, and interpersonal relationships.

Energy consumption at rate $a$ enables civilization to raise raw
materials across a potential energy barrier so that they can be incorporated
through diffusion into civilization's bulk at rate $j_{a}$. The amount
of energy that is required to turn raw materials into the stuff of
civilization is an enthalpy for rearranging matter into a new form.
Section \ref{sub:Systems-in-material-disequilibrium} included a discussion
of how heating transforms liquid into vapor within a pot of boiling
water. A similar ``phase transition'' might be seen when we burn
oil to extract such things as iron ore and trees from the ground.
Energy consumption continues as we reconfigure raw materials from
their low potential, natural state into carefully arranged steel girders
and houses where they becomes part of civilization's structure. 

In what we might call the economy, this energy consumption or heating
sustains all of civilization's existing internal circulations against
the continuous dissipation of heat at rate $d$ and material decay
at rate $j_{d}$. Civilization radiates heat to space while we and
our physical infrastructure fall apart. If civilization consumes energy
at rate $a$, largely through the exothermic reaction of primary energy
reserves (e.g. through combustion and nuclear reactions), and it dissipates
energy at an equivalent rate $d$, then the size of civilization stays
fixed. But if there is a disequilibrium where consumption exceeds
dissipation, then a remnant of power is able to go towards incorporating
new raw materials at rate $j_{a}-j_{d}$. 

Civilization falls under a class of ``emergent systems'' because
the disequilibrium allows civilization to expand into new reserves
of raw materials and energy, leading to a positive feedback that accelerates
growth.  From Eq. \ref{eq:dnbrevedt}, growth rates are equivalent
to the expansion of a length density $\Lambda$ that is tied to the
system's accumulated bulk to a one third power $N_{S}^{1/3}$. This
suggests that growth rates can be thought of as a lengthening and
concentration of the networks that form civilization's fabric. From
Eq. \ref{eq:eta-past}, we can infer that civilization growth is promoted
by the following factors: that civilization has access to large reserves
of available energy $\Delta H_{R}=N_{R}\Delta\mu$; that the amount
of energy $e_{S}^{tot}$ that is required to incorporate raw materials
into civilization's structure is low; and that civilization does not
fray too quickly, such that the decay parameter $\delta=j_{d}/j_{a}$
expressing relative rates of decay is small%
\footnote{As a practical matter, $N_{R}$ might be expressed by civilization
in units of millions of barrels of oil equivalent (mmboe), where the
potential energy of combustion contained in one barrel is equivalent
to $\Delta\mu$. %
}. 

In what follows, these concepts are extended to provide specific formulations
for the long-term evolution of civilization, expressible in such purely
fiscal terms as rates of return on wealth, economic production, innovation,
and technological change.

\subsection{Expression of fiscal quantities in thermodynamic terms}

\begin{figure}[h]
\noindent \begin{centering}
\includegraphics[width=4.5in]{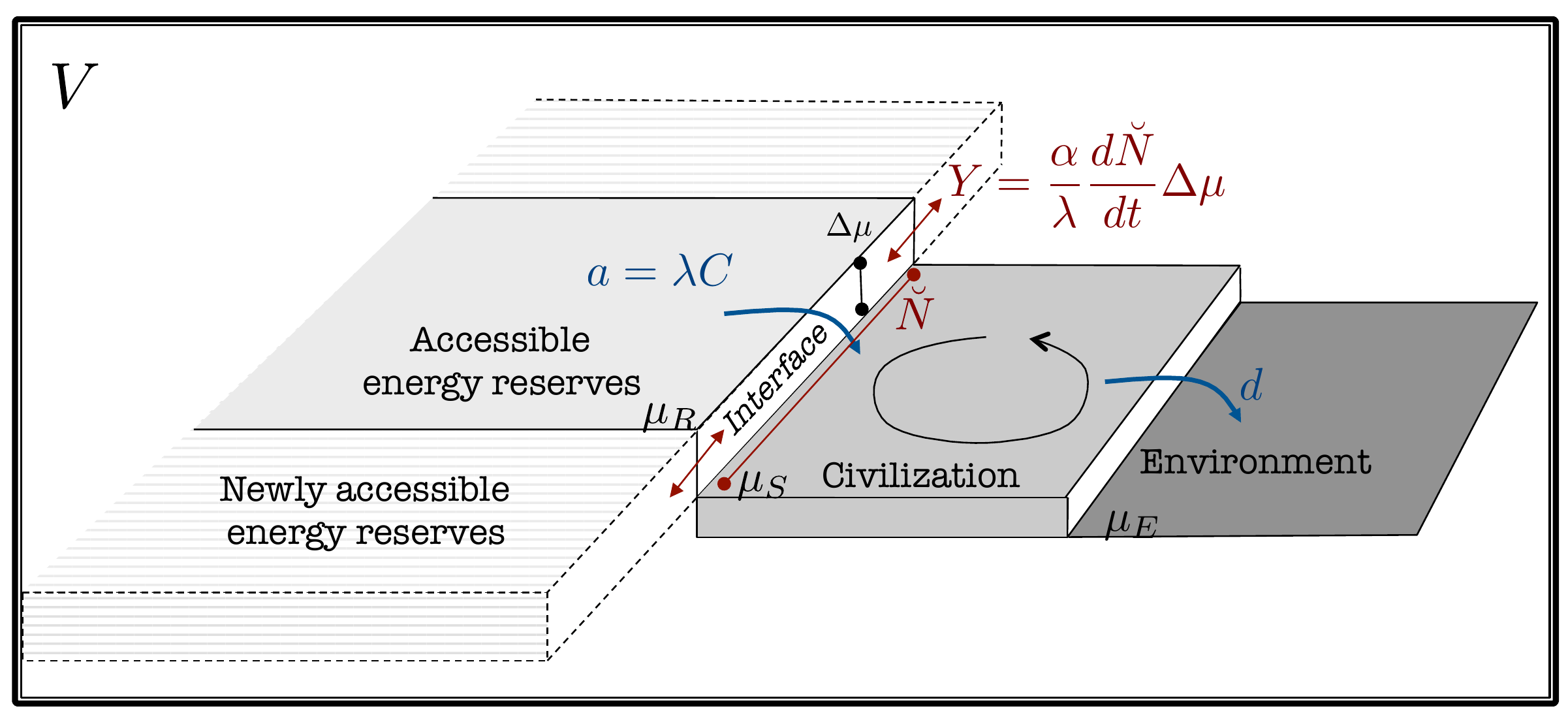}
\par\end{centering}

\caption{\label{fig:civilization}\emph{Representation of Fig. \ref{fig:diffusion}
in terms of global fiscal wealth $C$ and economic production $Y$,
as linked to rates of primary energy consumption $a$ and the size
of an interface with respect to energy reserves $\breve{N}$. Economic
production $Y$ is tied to interface growth, representing an expansion
of the capacity of civilization to draw from newly accessible energy
reserves. Energy consumption sustains civilization circulations against
dissipation to the environment at rate} $d$.}
\end{figure}
In \citet{GarrettCO2_2009}, it was hypothesized that global rates
of energy consumption $a$ are linked to a very general metric of
global economic wealth $C$ through a constant $\lambda$: 
\begin{equation}
a=\lambda C\label{eq:alambdaC-1}
\end{equation}
where current wealth is viewed as the time integral of past inflation-adjusted
economic production 
\begin{equation}
C=\int_{0}^{t}Y\left(t'\right)dt'\label{eq:CintY}
\end{equation}
The motivation for these expressions was that global energy consumption
at rate $a$ sustains the internal circulations of civilization against
an associated power dissipation $d$. If the capacity to sustain the
global economy's circulations is what we implicitly value, then primary
energy consumption should be fundamentally tied to a general representation
of economic wealth (Fig. \ref{fig:civilization}).

The hypothesis that $\lambda$ is a constant is falsifiable. Since
Gross Domestic Product GDP is the total productivity within a period
of one year, Eq. \ref{eq:CintY} can be calculated from
\begin{equation}
C_{i}=\sum_{i}\mbox{GDP}_{i}\label{eq:C_GDP}
\end{equation}
where $i$ is a time index starting from the beginnings of civilization.
Historical estimates of world GDP are available from such sources
as \citet{Maddison2003} and can be used to calculate $C$ as outlined
in Appendix C of \citet{GarrettCO2_2009}. Combined with available
statistics for global primary energy consumption, Eq. \ref{eq:alambdaC-1}
was shown to be supported by the data. Expressed in inflation-adjusted
2005 US dollars, available statistics indicate that $\lambda$ has
a maintained a steady value for the past few decades for which global
statistics for $a$ are available. Effectively, what sustains the
purchasing power embodied in each one thousand dollar bill, and distinguishes
it from a mere piece of paper, is a continuous 7.1$\pm$ 0.1 Watts
of primary energy consumption. 

Alternatively, in the year 2009, a global wealth of 2290 trillion
U.S dollars was supported by 16.1 terawatts of primary energy consumption.
In 1980, 1303 trillion 2005 dollars was sustained by 9.6 terawatts.
In the interim, the ratio of these two quantities was essentially
unchanged \citep{GarrettCO2_2009,GarrettRMJ2012}. Thus, it appears
that fiscal wealth can be considered to be a human representation
of the magnitude of the associated circulations that power consumption
can support. 

While wealth, as defined by Eq. \ref{eq:alambdaC-1}, has units of
currency and therefore might appear to be much like the term ``capital''
used in traditional economic treatments \citep[e.g.,][]{Solow1956},
there is a key difference. The term capital is normally reserved for
the additive value of fixed ``physical'' structures such as buildings
and roads. Economic output $Y$ is not considered to be directly additive
to physical capital because a portion is ``consumed'' by people
rather than ``saved'' for the future. The motivation for this approach
is that it seems logical to focus on people apart from non-living
structures given that, after all, the economy is human; human labor
uses physical capital to enable future consumption and certainly not
the reverse. 

This traditional approach offers a self-consistent way to track financial
accounts, but it approaches economic growth as if it does not need
to directly acknowledge universal physical laws. A lack of appeal
to resource constraints has been pointed out by many others \citep[e.g.,][]{Georgescu-Roegen,Costanza1980,WarrAyres2006}.
However, one point that has been missed is that the Second Law of
Thermodynamics forbids the existence of isolated systems, either in
space or time, and this places constraints on what an economic growth
model should look like \citep{GarrettRMJ2012}. 

For example, where the human and physical components of civilization
are interconnected, they cannot be mathematically treated as being
independent. This means that labor cannot be easily separated from
physical capital, and physical capital cannot be treated as being
purely additive. All aspects of civilization are intertwined through
their networks. People need houses as much as houses need people in
order to maintain their respective worth; removing one affects the
worth of the other. In the same vein, human consumption cannot disappear
to the past because the past is intertwined with the present. Even
if someone is only ``consuming'' a hamburger, a hamburger is nourishing
and satisfying in a way that both enables human interactions with
the rest of civilization and carries a memory of the pleasures of
hamburger consumption into the future. Even for ourselves, the thoughts
in our brain cannot be meaningfully separated from our cardiovascular
system and stomachs; each has no independent value since as each needs
the others to work. 

So there is no embodied value within any object itself, but only within
its ties to other elements of civilization. A brick of solid gold
is worth nothing if it is forgotten and lost in the middle of the
desert, but much more if it facilitates financial flows through its
integration within an economic network. Wealth includes people, their
knowledge, their buildings, and their roads, but only insofar as they
are interconnected through networks to the rest of the whole. The
elements of networks cannot easily be treated as being mathematically
additive, as in traditional economic treatments. Rather their value
is only in how their relationships facilitate the internal circulations
that demand civilization-scale thermodynamic flows. 

So, here the approach is to treat civilization as a system with constant
potential $\mu_{S}$ (Fig. \ref{fig:civilization}) whose collective
wealth is a fiscal expression of how its elements are intertwined
in a way that mutually supports global scale diffusive and dissipative
flows. From Eqs. \ref{eq:a_alphanhat} and \ref{eq:alambdaC-1}, 
\begin{equation}
C=\frac{\alpha}{\lambda}\breve{N}\Delta\mu\label{eq:CNbreve}
\end{equation}
where through Eq. \ref{eq:Nbreve}, $\breve{N}$ is related to the
system size through $N_{S}^{1/3}$ and a quantity of potential energy
$N_{R}\Delta\mu$. Or, from Eq. \ref{eq:aflowslambda}, 
\begin{equation}
C=\frac{\mathcal{D}}{\lambda}\Lambda\Delta H_{R}\label{eq:C_Lambda}
\end{equation}
The financial value of civilization lies in the total length density
of a global network $\Lambda$, with the caveat that the total network
must be coupled to reserves of potential energy $\Delta H_{R}$ that
enable diffusive flows with diffusivity $\mathcal{D}$. Expressing
the diffusion of knowledge and goods within human systems in terms
of a network length density and a proximity to resources is in fact
a common approach to human systems, albeit one that is normally discussed
in less strictly thermodynamic terms \citep[e.g.,][]{barabasi1999emergence,jackson2010social,bahar2012international}
.

The complexity of civilization is extraordinary, and it would be extremely
challenging if not impossible to model all possible interactions within
the network. While nothing forbids looking at civilization's internal
components alone, as a first step, thermodynamic principles offer
simplification of lowering resolution so that human and physical capital
are regarded at global scales. With this approach, the trade-off is
that nothing can be said about the internal details of civilization,
except perhaps in a statistical sense \citep[e.g.,][]{Ferrero2004}.
The advantage is that it enables a straight-forward link between physical
and fiscal quantities. Stepping back to view civilization as a whole
simplifies the relevant economic growth equations by removing the
complexities of internal communications and trade.

\subsection{Thermodynamics of nominal and inflation-adjusted economic production}

Where fiscal wealth is defined holistically by $C=\int_{0}^{t}Y\left(t'\right)dt'$
(Eq. \ref{eq:CintY}), there appears to be a fixed relationship to
thermodynamic flows through $a=\lambda C$ (Eq. \ref{eq:alambdaC-1}).
Thus, the very general physical principles derived in Section \ref{sec:Energetic-and-material-flows}
can now be applied to derive economic production functions that are
expressible in units of currency.

The simplest expression of the production function is that it adds
to economic wealth as it has been defined above: 
\begin{equation}
Y=\frac{dC}{dt}\label{eq:dCdtY}
\end{equation}
where $Y$ is inflation-adjusted (or real) economic output or productivity,
with units of currency per time. However, since $a=\lambda C$, $w=\Delta\mu d\breve{N}/dt$,
and from Eqs. \ref{eq:CNbreve} and \ref{eq:C_Lambda}, any of the
following expressions also apply, where $\alpha$, $\mathcal{D}$
and $\lambda$ are constants: 
\begin{eqnarray}
Y & = & \frac{1}{\lambda}\frac{da}{dt}\label{eq:GDPwork}\\
 & = & \frac{\alpha}{\lambda}w\\
 & = & \frac{\alpha}{\lambda}\frac{d\breve{N}}{dt}\Delta\mu\\
 & = & \frac{\mathcal{D}}{\lambda}\frac{d}{dt}\left(\Lambda\Delta H_{R}\right)
\end{eqnarray}
Perhaps rather intuitively, economic production is directly tied to
the amount of physical work $w$ that is done to expand the capacity
to consume energy through an increase in network density $\Lambda$
and an expansion of available energy reserves $\Delta H_{R}$. Real
production is valuable to the extent that it accelerates the energetic
flows $a$ that sustain civilization's circulations. From Eq. \ref{eq:CintY}
current global wealth is a consequence of past net work $C=\left(\alpha/\lambda\right)\int_{0}^{t}wdt'$.
From Eq. \ref{eq:work}, net work $w$ expands the material interface
$\breve{N}$ between civilization and the primary energy reserves
that sustain it (Fig. \ref{fig:civilization}). Net work is done where
there is an imbalance between consumption and dissipation, allowing
civilization to incorporate matter into its structure faster than
it decays.

From Eqs. \ref{eq:eta_logarithm} and \ref{eq:alambdaC-1}, a more
purely fiscal expression of the production function is one that is
related to wealth and rates of energy consumption through 
\begin{equation}
Y=\frac{dC}{dt}=\eta C\label{eq:YetaC}
\end{equation}
where, $\eta$ is the rate of emergent growth for thermodynamic systems.
For economic systems, the rate of emergent growth $\eta$ can be termed
more fiscally as the ``rate of return'' since, like money in the
bank, the rate of return expresses the growth rate of global wealth
through 
\begin{equation}
\eta=\frac{d\ln C}{dt}\label{eq:etalnC}
\end{equation}

In \citet{Garrettcoupled2011}, it was argued that this rate of return
$\eta$ can be expressed in terms of two components $\eta=\beta-\gamma$,
expressing a source and a sink, in which case production is related
to wealth through 
\begin{eqnarray}
Y & = & \left(\beta-\gamma\right)C\nonumber \\
 & = & \hat{Y}-\gamma C\label{eq:nominal-production}
\end{eqnarray}
where $\beta$ is a coefficient of nominal production, $\hat{Y}=\beta C$
is the nominal economic output, and $\gamma C$ is the magnitude of
any correction to nominal production that is required to yield inflation-adjusted
real production. From Eq. \ref{eq:eta_approx}, this implies a link
to the rates of material consumption and decay through, 
\begin{eqnarray}
\beta & = & \frac{j_{a}}{3N_{S}}\label{eq:beta}
\end{eqnarray}
 and
\begin{eqnarray}
\gamma & = & \frac{j_{d}}{3N_{S}}\label{eq:gamma}
\end{eqnarray}
 or, from Eq. \ref{eq:delta} 
\begin{equation}
\gamma=\delta\frac{j_{a}}{3N_{S}}\label{eq:gamma-1}
\end{equation}
Expressed thermodynamically, $\beta$ can be viewed as a rate coefficient
for growth and $\gamma$ as a rate coefficient for decay, each with
units of inverse time. 

Normally, the the GDP deflator is what is used to represent the degree
of any revisions to calculations of nominal output, i.e., the nominal
GDP is revised downward by a factor $\hat{Y}/Y$ . The GDP deflator
is linked to inflation insofar that it is estimated from price changes
in a very broad, moving basket of goods. For inter-annual calculations,
the factor by which the nominal GDP must be adjusted to be compared
to the nominal GDP in a prior year is:
\begin{equation}
\mathrm{GDP\, Deflator}=\frac{\hat{Y}}{Y}\simeq1+\left\langle i\right\rangle \label{eq:GDP_deflator}
\end{equation}
where $\left\langle i\right\rangle $ is the calculated average inflation
rate for the year. Assuming the inflation rate is much less than 100\%
per year, it follows that 
\begin{equation}
\left\langle i\right\rangle =\frac{\hat{\mathrm{GDP}}-\mathrm{GDP}}{\hat{\mathrm{GDP}}}\simeq\frac{\hat{Y}-Y}{\hat{Y}}=\frac{\left\langle \gamma\right\rangle }{\left\langle \beta\right\rangle }\label{eq:inflation_gamma-beta}
\end{equation}
From Eqs. \ref{eq:beta} and \ref{eq:gamma-1}, this leads to the
very simple result that global-scale inflation rates can be viewed
as a fiscal expression of the decay parameter $\delta=j_{d}/j_{a}$:
\begin{eqnarray}
\left\langle i\right\rangle  & = & \frac{\left\langle \gamma\right\rangle }{\left\langle \beta\right\rangle }\label{eq:inflation-delta}\\
 & \simeq & \left\langle \delta\right\rangle =\frac{\left\langle j_{d}\right\rangle }{\left\langle j_{a}\right\rangle }\nonumber 
\end{eqnarray}

The interpretation might be that civilization decay is an inflationary
pressure on economic production because it ``devalues'' the productive
capacity of existing assets by taking away that which has previously
been built, learned, or born. This fraying of networks occurs because
people die or forget, buildings crumble, and machines oxidize. For
example, it has been estimated that 10\% of our twentieth century
accumulation of steel has been lost to rust and war \citep{Smil2006}.
Where human and physical networks fall apart, there is a diminished
capacity to enable the thermodynamic flows that sustain civilization
wealth. Any monetary assets that were previously created to support
human and physical wealth no longer possess the same real purchasing
power.%
\footnote{Deflation (or negative inflation) is associated with $\left\langle i\right\rangle \simeq\left\langle \delta\right\rangle <0$,
which can be satisfied provided that $j_{a}<0$. Negative raw material
consumption might arise where raw materials are sourced from within
rather than without. %
}

To see the sources of inflationary trends, since $j_{a}=a/e_{S}^{tot}$
(Eq. \ref{eq:ja}) and $a\propto\Delta H_{R}$ (Eq. \ref{eq:flowsNbreve_general}),
then assuming that $e_{S}^{tot}$ changes slowly:
\begin{eqnarray}
\frac{d\ln\left\langle i\right\rangle }{dt} & = & \frac{d\ln\left\langle j_{d}\right\rangle }{dt}-\frac{d\ln\left\langle j_{a}\right\rangle }{dt}\label{eq:inflation-changes}\\
 & \simeq & \frac{d\ln\left\langle j_{d}\right\rangle }{dt}-\frac{d\ln\left\langle \Delta H_{R}\right\rangle }{dt}\nonumber 
\end{eqnarray}
So, rising inflation might occur if material decay $j_{d}$ accelerates,
perhaps from the types of global scale natural disasters that might
be associated with climate change \citep{Zhang_war_climate,Lobell2011}.
Alternatively, inflation might be driven by a declining availability
of energy reserves $\Delta H_{R}$ \citep{Bernanke1997}. 

As a caution, traditional interpretations of price inflation \citep[e.g.,][]{Parkin2008}
may not be a perfect match for the treatment described here. Pure
price inflation is a form of devaluation that arises because existing
monetary wealth has a lower purchasing power, so it is often viewed
as being simply a matter for control by central banks. 

However, the very general expression of wealth $C$ that has been
discussed here extends beyond money and physical assets to comprise
our physical and human relationships. In this case, devaluation might
arise because previously acquired skills might no longer be needed
by others because our capacity for work goes idle for lack of an energetic
impetus. Car production might decline if oil becomes scarce and expensive.
The workers and their factories remain but the external demand for
petroleum driven transportation declines and this leads to car manufacturer
layoffs \citep{lee2002dynamic}. Unemployment is just another side
of a more general inflationary coin%
\footnote{In fact, and apparent short-term trade-off between unemployment and
price inflation is well known in the field of Economics and has been
termed the ``Phillips Curve'' \citep{Phillips1958}. %
}. 

Of course, governments might rebuild human networks through financial
investments that bring workers back into paying jobs. But to have
a sustained effect on economic output, the hope would need to be that
these investments lead to a commensurate increase in energetic consumption
(Eq. \ref{eq:GDPwork}). Real civilization wealth and energy consumption
are intertwined through $a=\lambda C$ (Eq. \ref{eq:alambdaC-1}),
where wealth is tied to a capacity to access resources (Eq. \ref{eq:CNbreve}).
Simply printing money does not add to real wealth; being able to access
new energy reservoirs does. Stimulating the economy by loosening the
availability of money may be associated with nominal production in
the short term (Eq. \ref{eq:nominal-production}); but, if it fails
to ultimately create or be associated with a sustained increase in
energy consumption, the thermodynamics suggests that there will be
an offset to nominal wealth production through some combination of
unemployment and price inflation (Eq. \ref{eq:inflation_gamma-beta}).

\section{Thermodynamics of technological change, innovation, and growth}

Thus far, it has been shown that an economic growth model can be defined
by the coupled equations for the production function for real output
$Y$, and the growth of real wealth $C$ given by $dC/dt=Y$ and $Y=\eta C$,
where $\eta$ is a variable rate of return on wealth. As described
in \citet{GarrettCO2_2009}, these equations can be viewed as being
a more thermodynamically based (and dimensionally self-consistent)
form of the Solow-Swan neo-classical economic growth model \citep{Solow1956},
where $C$ is a generalized form of physical capital ($K$) that encompasses
labor ($L$), and $\eta$ is analogous to the total factor productivity
($A$), whose changes relate to technological change ($d\ln A/dt$). 

Technological change is often seen as a primary driver of long-run
economic growth \citep{Solow1957} but the source of technological
change remains somewhat of a puzzle. Sometimes it is regarded as having
endogenous origins, perhaps due to government investments in research
and development \citep{Romer1994}. However, the forces behind technological
change can also be seen in light of a more strictly thermodynamic
context. The rate of return $\eta$ evolves according to a deterministic
expression obtained by taking the derivative of the logarithm of Eq.
\ref{eq:eta-past}:

\begin{eqnarray}
\frac{d\ln\eta}{dt} & = & -2\frac{dN_{S}/dt}{3N_{S}}+\frac{d\ln\mbox{\ensuremath{\left(1-\delta\right)}}}{dt}+\frac{d\ln\Delta H_{R}}{dt}-\frac{d\ln e_{S}^{tot}}{dt}\label{eq:dlnetadt_conductivity}\\
 & = & -2\eta+\eta_{\delta}+\eta_{R}^{net}-\eta_{e}\nonumber \\
 & = & -2\eta+\eta_{tech}
\end{eqnarray}
Here, the term $d\ln\eta/dt$ is referred to as \emph{economic innovation}
because positive values of Eq. \ref{eq:dlnetadt_conductivity} represent
an acceleration of existing rates of return $\eta$. Innovations are
what are required for rates of return on wealth to rise. Defining
$\tau_{\eta}=1/\left(d\ln\eta/dt\right)$ as the characteristic time
for innovation, then wealth grows from an initial value $C_{0}$ as
\begin{equation}
C=C_{0}e^{\eta\tau_{\eta}\left(e^{t/\tau_{\eta}}-1\right)}\label{eq:super-exponential}
\end{equation}
If innovation is positive, then wealth grows explosively or super-exponentially.
In the limit of no innovation and $\tau_{\eta}\rightarrow\infty$,
the growth of wealth reduces to the simple exponential form $C=C_{0}\exp\left(\eta t\right)$
\citep{GarrettCO2_2009}.

The sum $\eta_{tech}=\eta_{\delta}+\eta_{R}^{net}-\eta_{e}$ is termed
here as the \emph{rate of technological change} $\eta_{tech}$ because
it is the driving force behind innovation $d\ln\eta/dt$. It represents
the sum of reductions to net decay ($\eta_{\delta}$), rates of net
energy reserve expansion ($\eta_{R}^{net}$), and reductions in the
amount of energy required to access raw materials (-$\eta_{e})$.
The following examines each component of technological change in more
detail.

\subsection{Innovation through increased longevity}

The first component of technological change is $\eta_{\delta}$, which
relates to reductions in the decay parameter $\delta$ (Eq. \ref{eq:delta}).
From Eq. \ref{eq:inflation-delta}, and assuming that the global inflation
rate is much less than 100\%, the decay parameter is approximately
equal to the inflation rate through $\left\langle i\right\rangle \simeq\left\langle \delta\right\rangle $.
In this case, the first order expansion in $\eta_{\delta}$ yields
\begin{equation}
\eta_{\delta}=\frac{d\ln\left(1-\delta\right)}{dt}\simeq-\frac{d\left\langle \delta\right\rangle }{dt}\label{eq:eta-delta}
\end{equation}
Since $\delta=j_{d}/j_{a}$ (Eq. \ref{eq:delta}), one way of interpreting
$\eta_{\delta}$ is through 
\begin{equation}
\eta_{\delta}=-\frac{1}{j_{a}}\left(\frac{\partial j_{d}}{\partial t}\right)_{j_{a}}\label{eq:eta-delta-ja}
\end{equation}
or, for a given rate of material consumption $j_{a}$, innovation
is favored by decreasing decay rates $j_{d}$. If people are enabled
to live longer through advancements in health \citep{HealthandEconomics2005},
or their structures are built so that they last longer \citep{kalaitzidakis2004macroeconomic},
then this is a form of positive technological change that contributes
to faster growth. Since the decay parameter is related to the inflation
rate, it follows that this innovative force would show up in global
scale economic statistics as declining inflation. In other words:
\begin{equation}
\eta_{\delta}\simeq\frac{d\left\langle \delta\right\rangle }{dt}\simeq-\frac{d\left\langle i\right\rangle }{dt}\label{eq:eta-delta-inflation}
\end{equation}

\subsection{Innovation through discovery of energy reserves}

The expression $\eta_{R}^{net}$ refers to the net rate of expansion
of available energy reserves $\Delta H_{R}$. Having a newly plentiful
supply of energy accelerates economic innovation and growth \citep{Smil2006,AyresWarr2009}. 

There are two forces here. One is for energy reserves to decline due
to potential energy consumption at rate $a$. The second is that civilization
discovers new reserves of energy at rate $D$. The balance of these
two forces is given by 
\begin{eqnarray}
\frac{d\ln\Delta H_{R}}{dt} & = & \frac{\mathrm{Discovery}-\mathrm{Depletion}}{\mathrm{Existing}}\label{eq:discovery}\\
 & = & \frac{D-a}{\Delta H_{R}}\nonumber \\
 & = & \eta_{D}-\eta_{R}\nonumber 
\end{eqnarray}
Net reserve expansion occurs when rates of reserve discovery $\eta_{D}$
exceed rates of reserve depletion $\eta_{R}$, requiring that $\eta_{D}/\eta_{R}>1$. 

As illustrated in Fig. \ref{fig:civilization}, civilization consumes
energy as it grows, and it grows into surroundings that may or may
not contain new reserves of fuel \citep{Murphy2010}. If $\eta_{D}/\eta_{R}>1$,
then civilization discovers new reserves faster than it depletes previously
discovered reserves. In some global sense, energy becomes ``cheaper''
relative to the existing quantity of wealth $C$, allowing the rate
of return on wealth $\eta$ to be higher than it would be otherwise.

\subsection{Innovation through increased efficiency of raw material extraction}

The expression $-\eta_{e}$ in Eq. \ref{eq:dlnetadt_conductivity}
refers to changes in the specific enthalpy of civilization $e_{S}^{tot}$.
Since $e_{S}^{tot}=a/j_{a}$ (Eq. \ref{eq:ja}), a decline in $e_{S}^{tot}$
would appear as a decrease in the amount of power $a$ that is required
for civilization to extract raw materials and incorporate them into
civilization at rate $j_{a}$. 

Comparing Eqs. \ref{eq:dlnLambda/dt} and \ref{eq:eta_approx}, civilization
networks grow through raw material consumption. If growing civilization
requires less energy per unit matter, then civilization can grow faster
for any given rate of global energy consumption $a$. Since $a=\lambda C$,
the implication is that raw materials have become cheaper relative
to total global wealth $C$. This is an innovative force because the
material growth of civilization increases access to the resources
that sustain it.

\subsection{\label{sub:Innovation-through-higher-connectivity}Innovation through
a common culure}

I suggest that a second interpretation of the expression $-\eta_{e}$
in Eq. \ref{eq:dlnetadt_conductivity} is that economic innovation
can be derived from seeking a common culture. The specific enthalpy
of civilization can be expressed as $e_{S}^{tot}=\nu e_{S}$ (Eq.
\ref{eq:e_s_tot_nu}), where $e_{S}$ is the specific energy of each
independent mode and $\nu$ represents the number of orthogonal (or
independent) modes within a mechanical system. The energy associated
with each mode $e_{S}$ can be assumed to be equal through the equipartition
principle, provided sufficiently long timescales are considered. 

For the purpose of facilitating the thermodynamics in this treatment,
civilization is considered to lie along a surface of constant potential,
in which case $e_{S}$ does not change. However, the number of degrees
of freedom $\nu$ in the system remains a free parameter. If innovations
occur when $e_{S}^{tot}$ declines, this is due to a decrease in $\nu$. 

For guidance on what this rather abstract thermodynamic result actually
means, it might help to first look at the behavior of a ``stiff''
molecule such as gaseous molecular nitrogen (N$_{2}$). Nitrogen can
be idealized as two nitrogen atoms connected by a stiff spring. At
room temperatures, N$_{2}$ has five orthogonal degrees of freedom
that determine its specific enthalpy $e_{S}^{tot}$. Three of these
come from molecular translational motions within the three dimensions
of space; two come from orthogonal rotational motions. Two additional
degrees are added to account for molecular pressure to yield $\nu=7$.
If the specific energy (or temperature) increases ten-fold, vibrational
transitions within N$_{2}$ no longer stay ``frozen out'', and $\nu$
increases from seven to nine. 

So, molecules that are more internally ``stiff'' have a smaller
number of independent degrees of freedom for molecular motion. At
room temperatures, N$_{2}$ has relatively low values of $e_{S}=kT/2$
that prevent the individual atoms from oscillating independently.
The stiffness of the bond requires that the two nitrogen atoms rotate
and translate \emph{together,} as if they were connected.

With regards to civilization, we have witnessed an extraordinary increase
in internal connectivity through ever improving transport and communications
networks \citep{Dijk2012}. A way to interpret this growth in network
density is that it corresponds to a reduction in the effective number
of degrees of freedom in society. Technological change gives us a
more collective experience and global culture. 

For example, international trade allows us to consume very similar
products; transportation in the form of petroleum fueled cars is now
ubiquitous; and, we have now accepted English as our global \emph{lingua
franca}. Through communications, travel, international markets and
shipping, our world has become more interconnected and ``stiff''. 

There are some obvious tradeoffs to cultural similarity, setting aside
that a more homogeneous world is less interesting. Economic growth
and volatility is now sensed more globally. On one hand, civilization
might be fragile if it becomes too uniformly reliant on the same things.
A potato monoculture became susceptible to a blight that trade had
brought in the from the New World, leading to a catastrophic population
collapse known as the ``Irish Famine'' \citep{DonnellyJr.2001}.
A modern parallel is the proposition that our reliance on oil will
lead to a dramatic slowing of global economic growth should it rapidly
become scarce \citep{lee2002dynamic,Bardi2009,Sorrell2010,Murray481}.

On the other hand, increasing global common modes of transportation,
communication, and language facilitate innovation and growth. We build
roads for a reason, because they bring us together. By lowering the
specific enthalpy $e_{S}^{tot}=\nu e_{S}=a/j_{a}$ and reducing the
effective number of degrees of freedom $\nu$, less energy is required
to diffuse an equivalent quantity of matter through the structure
when new resources are uncovered.

\subsection{Diminishing returns as a drag on innovation}

The final term in Eq. \ref{eq:dlnetadt_conductivity}, $-2\eta$,
expresses a drag on how fast rates of return can grow. Innovation
naturally slows due to a \emph{law of diminishing returns}. In the
absence of innovation, wealth converges on a steady-state where rates
of return equal zero \citep{romer1986increasing}. Diminishing returns
exists as a force because current growth unavoidably becomes diluted
within the accumulated bulk that was built from past growth (Eq. \ref{eq:eta-past}).
Each incremental incorporation of raw materials into civilization
$j^{net}$ has a decreasing impact relative to the summation of previously
incorporated matter $\int_{0}^{t}j^{net}\left(t'\right)dt'$ . Diminishing
returns makes innovation rates increasingly negative and can only
be overcome if technological changes are sufficiently rapid. From
Eq. \ref{eq:dlnetadt_conductivity}, what is required is that $\eta_{tech}>2\eta$.

\section{Modes of growth in economic systems}

\subsection{Technological change and rates of return}

Because the above expressions are prognostic, the implication is that
there exist deterministic solutions for how rates of return on civilization
wealth and energy consumption change with time. Previous work has
identified characteristic sigmoidal or logistic behavior in the effects
of technological change on economic growth: after overcoming a period
of initial resistance, technological changes rapidly accelerate growth,
followed ultimately by saturation \citep{Landes2003,Smil2006,Marchetti2012}.
Indeed, Eq. \ref{eq:dlnetadt_conductivity} can be expressed in the
form of the logistic equation 
\begin{equation}
\frac{d\eta}{dt}=\eta_{tech}\eta-2\eta^{2}\label{eq:logistic}
\end{equation}
If rates of technological change $\eta_{tech}$ are constant, then
the solution has the sigmoidal or ``S-curve'' form 
\begin{equation}
\eta\left(t\right)=\frac{G\eta_{0}}{1+\left(G-1\right)\exp\left(-\eta_{tech}t\right)}\label{eq:logistic_solution}
\end{equation}
where $\eta_{0}$ is the initial value for the rate of return, and
\begin{equation}
G=\frac{\eta_{tech}}{2\eta_{0}}\label{eq:G}
\end{equation}
represents a ``Growth Number'' \citep{Garrettmodes2012} that partitions
solutions for $\eta\left(t\right)$ into varying modes of growth summarized
in Table \ref{tab:Modes-of-growth}. 
\begin{table}[h]
{\footnotesize \caption{\label{tab:Modes-of-growth}Modes of growth in economic systems.  Diminishing returns (DR), Technological Change (TC), Technological Decline (TD).}
}{\footnotesize \par}

{\footnotesize }%
\begin{tabular}{c|c|c|c|c|c}
\hline 
 & {\small Innovation} & {\small DR and TC} & {\small DR and TD} & {\small Decay} & {\small Collapse}\tabularnewline
\hline 
{\footnotesize $ $Growth number }{\small $\eta_{tech}/2\eta_{0}$} & {\footnotesize $G>1$} & {\footnotesize $0<G<1$} & {\footnotesize $G<0$} & {\footnotesize $G>1$} & {\footnotesize $G<1$}\tabularnewline
{\footnotesize Initial rate of return} & {\footnotesize $\eta_{0}>0$} & {\footnotesize $\eta_{0}>0$} & {\footnotesize $\eta_{0}>0$} & {\footnotesize $\eta_{0}<0$} & {\footnotesize $\eta_{0}<0$}\tabularnewline
{\footnotesize Limiting rate of return} & {\footnotesize $\eta_{tech}/2$} & {\footnotesize $\eta_{tech}/2$} & {\footnotesize $0$} & {\footnotesize $0$} & {\footnotesize $-\infty$}\tabularnewline
\hline 
\end{tabular}
\end{table}

The four modes of growth that are available to civilization are innovation,
diminishing returns, decay, and collapse, depending on the value of
$G$ and the initial value of $\eta$. Innovation is characterized
by growing rates of return; diminishing returns is associated with
declining rates of return, either to a limit of $\eta_{tech}/2$ or
to zero. Where rates of return are initially negative, decay either
slows with time or it accelerates in a mode of collapse.

\begin{figure}[h]
\includegraphics[width=4.5in]{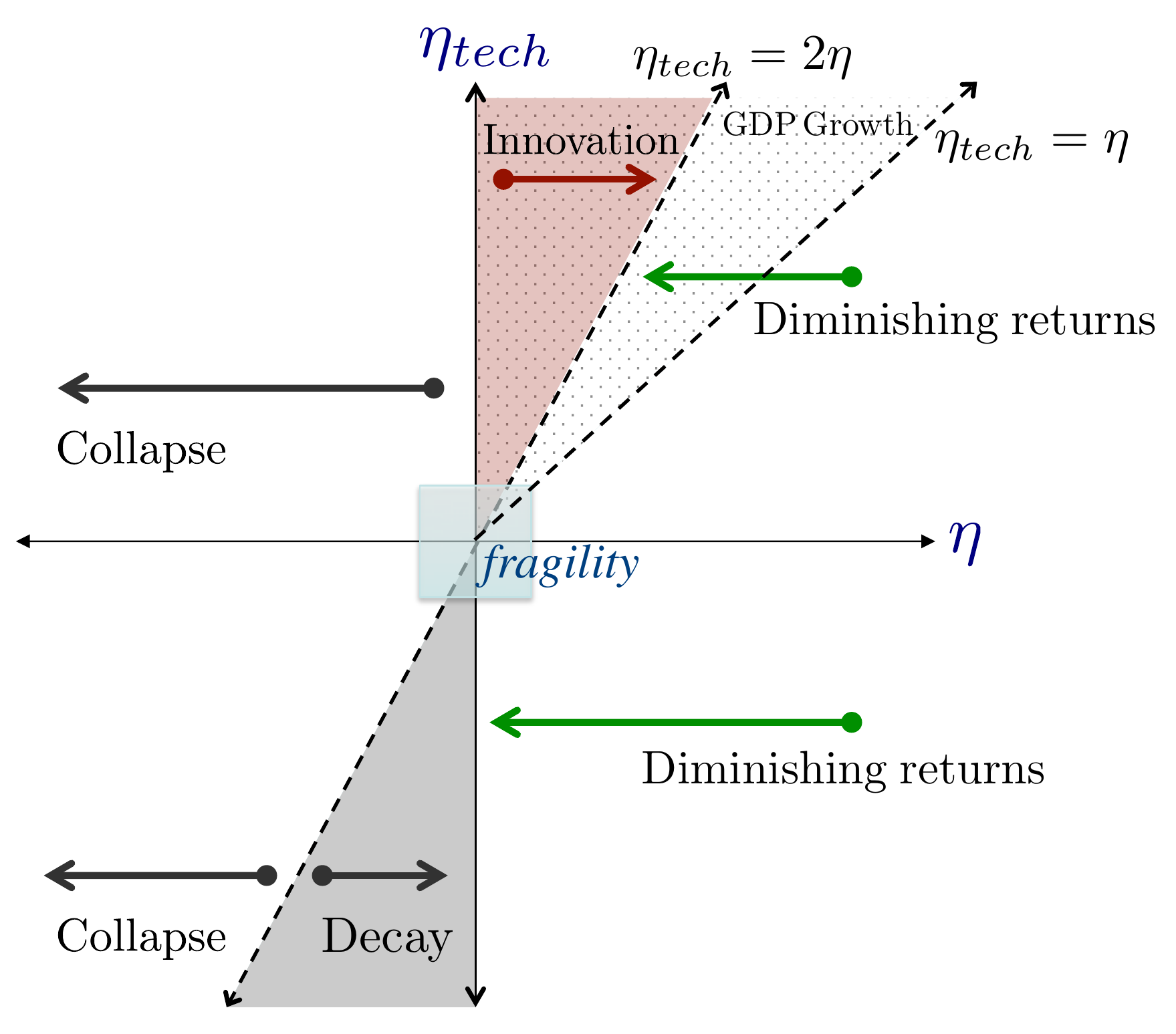}

\caption{\label{fig:innovation}Modes of growth in economic systems, partitioned
within a space of rates of technological change $\eta_{tech}$ and
rates of return on wealth $\eta$. Arrows represent trajectories for
rates of return, assuming that $\eta_{tech}$ is a constant. The dotted
region shows the domain of parameter space associated with GDP growth.
See text for details. }
\end{figure}
Fig. \ref{fig:innovation} carves these modes within a space of $\eta_{tech}$
and $\eta$, along with associated trajectories for any given value
of $\eta_{tech}$. For example, for values of $G>1$, civilization
is in a mode of innovation because technological innovation is sufficiently
rapid to overcome diminishing returns. At first, rates of return increase
exponentially but then they saturate to approach a value of $G\eta_{0}=\eta_{tech}/2$.
If $\eta$$ $ is initially 1 \% per year and rates of technological
change $\eta_{tech}$ are sustained at a nominal 4\% per year, then
one would expect rates of return $\eta$ to grow sigmoidally towards
2\% per year. The exponential phase of the sigmoidal growth would
have a characteristic time of $1/\eta_{tech}$, or 25 years.

\subsection{Technological change and GDP growth}

Innovation rates have a direct impact on rates of GDP growth. Since
$Y=\eta C$ (Eq. \ref{eq:YetaC-1}), and the rate of return is given
by $\eta=d\ln C/dt$ (Eq. \ref{eq:etalnC}), it follows that: 
\begin{equation}
\frac{d\ln Y}{dt}=\eta+\frac{d\ln\eta}{dt}\label{eq:GDPgrowth-1}
\end{equation}
GDP growth rates are a simple sum of the current rate of return $\eta$
and the innovation rate $d\ln\eta/dt$. GDP growth increases when
there is innovation.  

From Eq. \ref{eq:dlnetadt_conductivity}, the rate of return itself
evolves at rate $d\ln\eta/dt=-2\eta+\eta_{tech}$, where rates of
technological change $\eta_{tech}=\eta_{\delta}+\eta_{R}^{net}-\eta_{e}$
are a summation of reductions to net decay, net energy reserve expansion,
and improvements to the efficiency of raw material extraction and
incorporation into civilization. Since GDP growth is related to the
sum of the innovation rate and the rate of return in Eq. \ref{eq:GDPgrowth-1},
it follows that: 
\begin{equation}
\frac{d\ln Y}{dt}=-\eta+\eta_{tech}\label{eq:GDPgrowth_conductivity}
\end{equation}
The GDP growth rate is buoyed by positive technological change. However,
as illustrated in Fig. \ref{fig:innovation}, sustaining a growing
GDP in the long-term requires that: 
\begin{equation}
\eta_{tech}>\eta\label{eq:dlnsigmadt2}
\end{equation}
or that technological change must be more rapid than the current rate
at which energy consumption is growing $\eta=d\ln a/dt$. 

In fact, this poses an interesting quandary for economic growth. Supposing
that technological change is driven by net discovery of new energy
reserves (Eq. \ref{eq:discovery}), Eq. \ref{eq:dlnsigmadt2} implies
that sustaining GDP growth requires energy consumption to continue
to grow sufficiently fast that 
\begin{equation}
\frac{da}{dt}>\Delta H_{R}\left(Da-a^{2}\right)\label{eq:GDPgrowth_quandary}
\end{equation}
Eq. \ref{eq:GDPgrowth_quandary} is a logistic equation for energy
consumption \citep{Bardi2009,Hook2010}. Where energy consumption
rates $a$ are buoyed in the present by discovery of new reserves
at rate $D$, this acts as a drag on growth further down the road.
Other forms of technological change staying constant, the GDP approaches
a steady-state when consumption equals discovery and $a=D$.

\subsection{Fragility and growth}

How do civilizations ultimately decay and collapse? Obviously, rates
of return for civilization wealth must initially be positive for civilization
to have emerged in the first place. But positive growth cannot be
sustained forever because civilization networks are always falling
apart to some degree. And, on a world with finite resources, we will
eventually lose the capacity to keep fixing them. 

However, there is no spontaneous mathematical transition between modes
of growth that is implied by Eq. \ref{eq:logistic_solution}; in the
limit of $t\rightarrow\infty$, rates of return $\eta$ either asymptotically
approach a constant value, or they tend towards collapse. So transitions
between modes must be forced by some external impetus. In our case,
this might come from a rapid increase in global scale natural disasters,
perhaps due to climate change. 

\begin{figure}[h]
\includegraphics[width=4.5in]{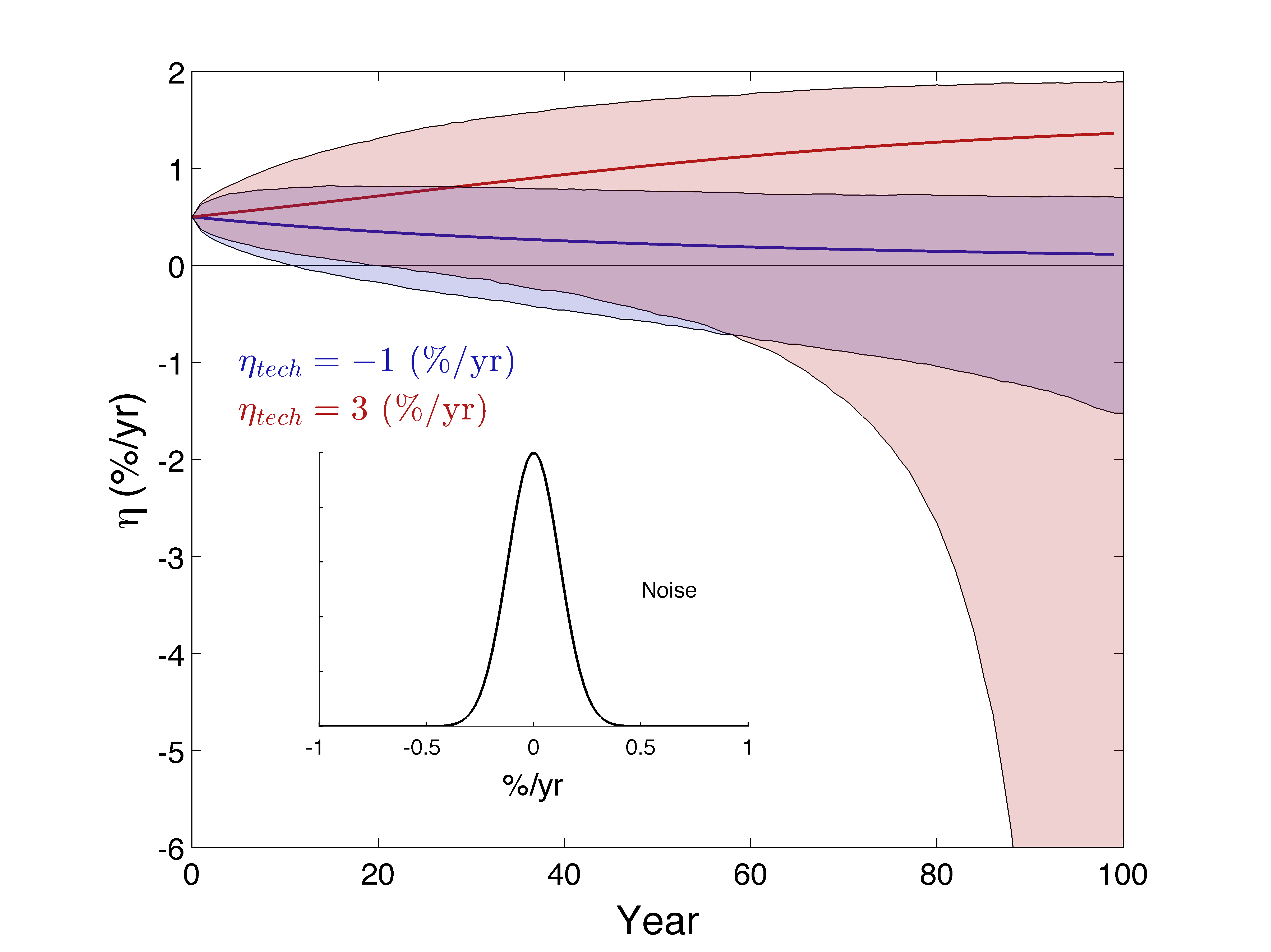}

\caption{\label{fig:future}For an initial value for the rate of return $\eta_{0}$
of 0.5\% per year, lines are trajectories of the evolution of $\eta\left(t\right)$
for scenarios with rates of technological change $\eta_{tech}$ of
3\% per year and -1\% per year, as given by Eq. \ref{eq:logistic_solution}.
The shaded region is derived from the upper and lower 5\% bounds in
an ensemble of 10,000 simulations where noise (inset) has been introduced
that has a standard deviation of 0.1\% per year for $\eta$. }
\end{figure}
For example, in Eq. \ref{eq:eta-past}, if the decay parameter $\delta=j_{d}/j_{a}$
is greater than unity, then $\eta$ must be negative. If material
decay exceeds material consumption, then civilization transitions
from positive to negative rates of return. Of course, things can go
both ways and Eq. \ref{eq:eta-past} also allows for conditions that
might suddenly favor growth, including decreased decay or significant
discoveries of new energy reserves that increase $\Delta H_{R}$.
These might permit civilization to transition from a mode of diminishing
returns into one of innovation and super-exponential growth.

\begin{figure}[h]
\includegraphics[width=4.5in]{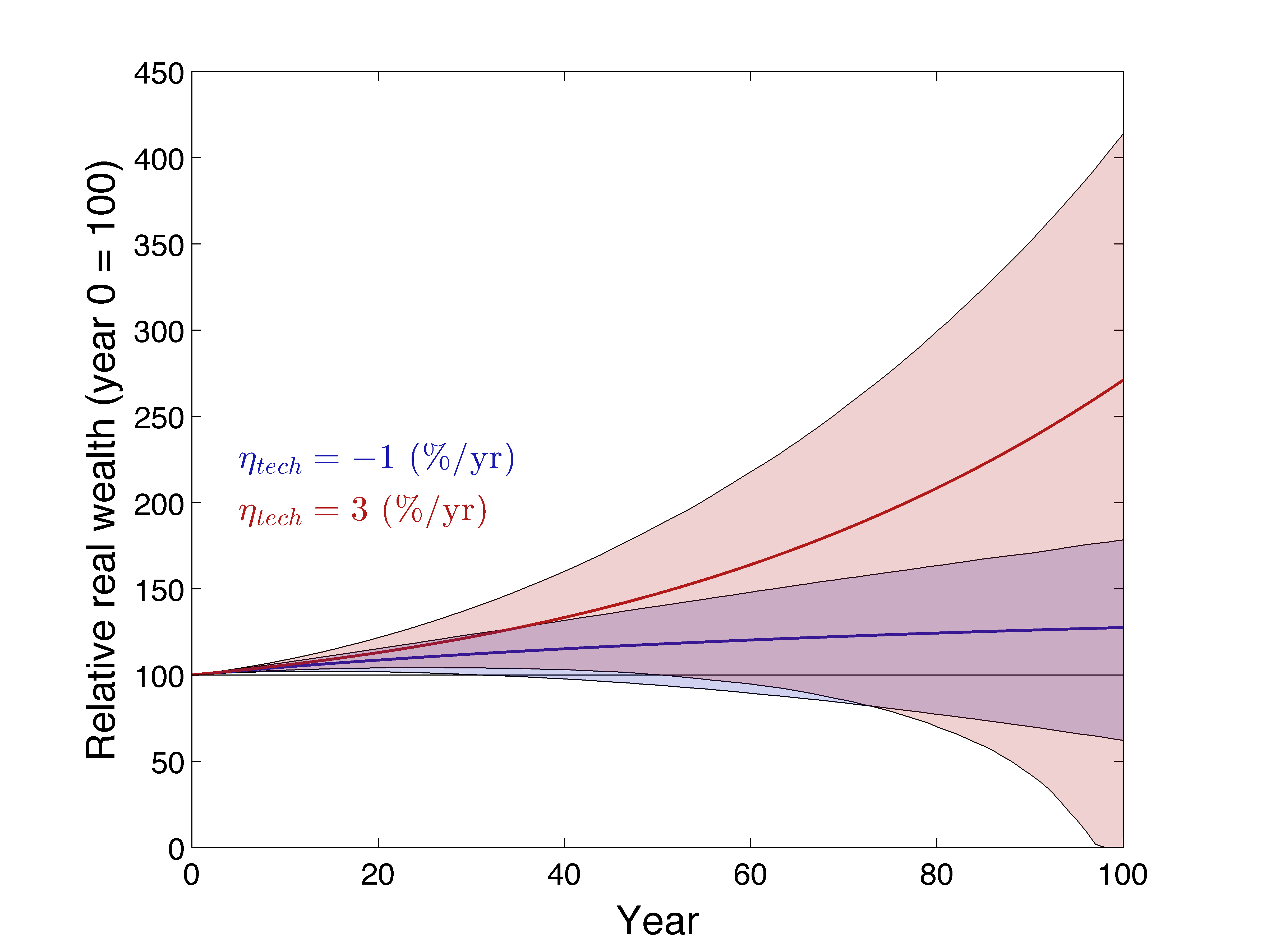}

\caption{\label{fig:future-wealth}For the scenarios shown in Fig. \ref{fig:future},
corresponding values of global inflation-adjusted wealth, referenced
to 100 in year 0.}
\end{figure}

To account for both the good and the bad in the future, stochastic
and largely unpredictable external events might be represented by
introducing noise to Eq. \ref{eq:eta-past}. An example of how this
might play out is illustrated in Figs. \ref{fig:future} and \ref{fig:future-wealth}.
If there is no noise, then trajectories follow the logistic solutions
provided by Eq. \ref{eq:logistic_solution}. But if random Gaussian
noise is added to $\eta$, then the range of possible trajectories
broadens. Notably, there are ``unlucky'' trajectories that could
be associated with frequent and persistent global scale natural disasters.
Disasters might push civilization into a transition towards a mode
of irreversible decay or collapse. Most notably, a transition is particularly
likely when rates of return approach zero. 

It has been pointed out that the existence of ``tipping points''
where there has been a ``slowing down'' is as a feature of ecological
and climate systems \citep{Dakos2008,Dakos2011}. What is interesting
in the simulations above is that the most dramatic rates of collapse
are associated with noisy trajectories that would otherwise be associated
with innovation and accelerating rates of growth. The same conditions
that allow for the human system to respond especially quickly to favorable
conditions are the some ones that allow the system to rapidly decay
quickly when conditions become unfavorable. Having a common culture
is a good example (Sec. \ref{sub:Innovation-through-higher-connectivity}).
It allows for exceptionally rapid diffusion of matter into civilization's
structure while also lending a fragility that permits co-ordinated
decline.

As illustrated in Fig. \ref{fig:innovation}, an innovative economy
that enjoys relatively rapid technological change with a growth number
$G>1$ might alternatively be viewed as a ``bubble economy'' that
lacks long-term resilience. Whether collapse comes sooner or later
depends on the quantity of energy reserves available to support continued
growth and the accumulated magnitude of externally imposed decay.
By contrast, an economy that is less innovative, with lower rates
of return $\eta$, has a lower risk of rapid rates of decline. In
the space shown in Fig. \ref{fig:innovation}, it lies ``farther
away'' from modes of collapse.

\section{Summary}

This paper has presented a physical basis for interpreting and forecasting
global civilization growth, by treating it as a thermodynamic system
that grows in response to interactions with its environment \citep{Garrettmodes2012}.
Like other living organisms  \citep{Vermeij2008}, civilization displays spontaneous emergent
behavior. Energy dissipation drives material
flows to civilization. If there is a net convergence of matter within
civilization, then civilization grows. Growth increases the availability
of new reserves and this leads to a positive feedback loop that allows
growth to persist. 

The negative feedback on growth is that civilization carries
with it a memory of its past. This slows growth through a ``law
of diminishing returns'' that is common to growing systems: current additions of matter become increasingly
diluted within an accumulation of past additions. Diminishing returns can
be overcome, but only if there is sufficiently rapid technological change. Technological
change has three broad categories: improved material longevity (or
reducing decay), the discovery of new reserves of energy, and increased
energy efficiency. One manifestation of higher energy efficiency might be a
 common global culture with fewer independent degrees of freedom,
because this decreases the amount of energy that is required
to diffuse raw materials throughout civilization's structure. 

These thermodynamic results can be expressed in purely fiscal terms
because there appears to be a fixed link between global rates of primary
energy consumption and a very general expression of human wealth:
$\lambda=$7.1$\pm$ 0.1 Watts of primary energy consumption is required
to sustain each one thousand dollars of civilization value, adjusting
for inflation to the year 2005 \citep{GarrettRMJ2012}. Wealth does
not rest in inert ``physical capital'', as in traditional treatments,
but rather in the density of connections between civilization elements,
insofar as this network contributes to a global scale consumption
and dissipation of energy (Eq. \ref{eq:C_Lambda}). 

The economic growth model for wealth $C$ and economic production
$Y$ is very simple: 
\begin{eqnarray}
\frac{dC}{dt} & = & Y\label{eq:growth_equations}\\
Y & = & \eta C\label{eq:YetaC-1}
\end{eqnarray}
where $\eta$ is a variable real rate of return on wealth, somewhat
analogous to the total factor productivity in traditional models.
The rate of return can be related to basic thermodynamic quantities
through 
\begin{eqnarray}
\eta & = & \alpha k\left(1-\delta\right)\frac{\Delta H_{R}}{N_{S}^{2/3}e_{S}^{tot}}\label{eq:rate-of-return-equations}
\end{eqnarray}
where $\delta$ relates civilization decay to how fast it incorporates
new raw materials, $\Delta H_{R}$ represents the quantity of available
energy reserves, $e_{S}^{tot}$ expresses the amount of energy required
to incorporate raw materials into civilization's structure, and $N_{S}=\int_{0}^{t}j^{net}dt'$
represents the accumulated size of civilization due to past raw material
flux convergence $j^{net}$. The constants $\alpha$ and $k$ are
unknown rate and shape coefficients; however, values of the rate of
return $\eta$ can be inferred from Eq. \ref{eq:YetaC-1}. For example,
current global rates of return are about 2.2 \% per year \citep{GarrettRMJ2012}.
What Eq. \ref{eq:rate-of-return-equations} shows is that trends in
$\eta$ can be forecast based on estimates of future decay and rates
of raw material and energy reserve discovery. 

Thus, Eqs. \ref{eq:growth_equations} through \ref{eq:rate-of-return-equations},
combined with the constant $\lambda$, offer a complete set of prognostic
expressions for civilization growth. The implications that have been
described are summarized as follows:
\begin{itemize}
\item Civilization inflation-adjusted wealth grows only as fast as rates
of global energy consumption.
\item Low inflation is maintained by high civilization longevity.
\item Rates of return on wealth decline when decay accelerates, or reserves
of raw materials and energy become increasingly scarce.
\item Through a law of diminishing returns, high current rates of return
imply a stronger drag on future growth. The mathematical form for
the evolution of rates of return is sigmoidal, as determined from
the logistic equation.
\item Rates of return grow when there is ``innovation''. As it is defined,
innovation is driven by technological change, but it must be sufficiently
fast to outweigh the law of diminishing returns. 
\item Global GDP growth requires energy consumption to grow super-exponentially,
or at an accelerating rate. GDP growth is sustainable for as long
as energy reserve discovery exceeds depletion. 
\item When growth rates slow and rates of return approach zero, civilization
becomes fragile with respect to externally forced decay. It lies along
a tipping point that might easily lead to a mode of accelerating decay
or collapse. 
\item Innovation and collapse are two sides of the same coin. Increased
internal connectivity allows for explosive growth when times are good,
but also for exceptionally fast decline when times turn bad. 
\end{itemize}
Many of these conclusions might seem intuitive, or as if they have
been expressed already by others from a more traditional economic
perspective. What is novel in this study is the expression of the
economic system within a deterministic thermodynamic framework where
a very wide variety of economic behaviors are derived from only a
bare minimum of ingredients. A sufficient set of statistics exists
for global economic productivity, inflation, energy consumption, raw
material extraction and energy reserve discovery that the model presented
here can be evaluated, and with no requirement for \emph{a priori }tuning
or fitting to historical data. If the analytical expressions are consistent
with past behavior, then this offers the possibility of providing
a range of physically constrained forecasts for future economic innovation
and growth. If not, then the model should be re-examined or discarded.

A follow-on paper will compare these prognostic formulations against
historical data. Civilization has enjoyed explosive growth since the
industrial revolution, but it is unclear how long this can be sustained
when it is facing ongoing resource depletion, pollution, and climate
change. The prognostic expressions that have been derived here will
be used to guide a physically plausible range of future timelines
for civilization growth and decay. 

\subsection*{Acknowledgments}
This work was supported by the Kauffman Foundation, whose views it does not claim to represent.



\end{document}